\documentclass{ws-ijmpd}
\usepackage[super,compress]{cite}
\usepackage[utf8]{inputenc}
\usepackage{graphicx,epsfig}  
\usepackage{subfigure}

\begin{document}


%
\catchline{}{}{}{}{}
%
\title{Astrometric microlensing}

\author{A. A. Nucita,  F. De Paolis, G. Ingrosso, M. Giordano, L. Manni}
\address{Department of Mathematics and Physics {\it `E. De Giorgi'}, University of Salento, Via per Arnesano, CP 193, I-73100,
Lecce, Italy}
\address{INFN, Sez. di Lecce, Via per Arnesano, CP 193, I-73100, Lecce, Italy}

\maketitle

\begin{history}
\received{Day Month Year}
\revised{Day Month Year}
\end{history}

\begin{abstract}
Astrometric microlensing will offer in the next future 
a new channel for investigating the nature of both lenses and sources involved in 
a gravitational microlensing event. The effect, corresponding to 
the shift of the position of the multiple image centroid with respect to the source star location, 
is expected to occurr on scales from  micro-arcoseconds to milli-arcoseconds depending on the characteristics of the lens-source system. Here, 
we consider different classes of events (single/binary lens acting on a single/binary source) 
also accounting for additional effects including the finite source size, the blending and orbital motion. This is particularly important in 
the era of Gaia observations which is making possible astrometric measurements with unprecedent quality.
\end{abstract}
\ccode{PACS numbers: 95.75.De, 97.10.Wn}

\section{Introduction}

Gravitational microlensing is a well known technique for detecting compact objects in the disk, bulge and  halo of our Galaxy\footnote{In the last years, a variant of 
the microlensing method, the so called pixel-lensing techinque demonstrated its capabilities to discover new microlensing events and variable stars also in the M31 galaxy (see, e.g., 
Refs. \refcite{nucita2007,nucita2009,nucita2010,nucita2014m31}).} via the amplification of the light of background sources. This possibility was offered by the technological advances in present 
instruments which allow to monitor (at the same time) millions of stars in large fields of view and to be sensitive to low-mass objects\cite{park2015} as well as to detect and characterize 
binary lens systems \cite{udalski2015} . In this respect, microlensing is also becoming a key method for discovering planetary systems 
with Earth-like planets orbiting their parent stars at distances of about a few AU and to obserserve free-floating planets \cite{sumi2011,hamolli2015} which, otherwise, would remain undetected.

An ongoing gravitational microlensing event also induces an astrometric shift between the light 
centroid of the multiple images and the source star position. This subject was studied by many authors 
(see e.g. Refs. \refcite{walker1995,miyamoto1995,hog1995,jeong1995,paczinsky1996,paczinsky1998,dominik2000,takahashi,lee2010}) who observed that in the simplest case of a point-like object lensing a 
single source the source image splits into two images with 
the position of the light centroid describing an ellipse with semi-axes depending from the lens impact parameter $u_0$ and the Einstein $t_E$.  When one considers a binary lens system 
\cite{han1999,griest1998,han2001,bozza2001,hideki,sajadian2015a}, the astrometric signal deviates from a ellipse being it strongly dependent 
on the binary system parameters (mass ratio and separation). The same happens when considering other effects as the blending (i.e., the fraction of light that does not get magnified but contributing 
to the photons collected during the observations), the finite-size source effect (related to the 
magnification of different parts of an extended source star) and the orbital motion of binary lenses and/or sources. Note that in all cases, the astrometric signal gives further information allowing one 
to alleviate the problem of the parameter degeneracy\footnote{Other methods that were recently considered rely on the measurement of the 
lens proper motion (see Ref. \refcite{propermotion}) and on polarization observations 
(see Refs. \refcite{ingrossoa,ingrossob}).} that afflicts the classical microlensing. 

The aim of the paper is to discuss the main observational features of astrometric microlensing and introduce second order effects that may be detcatable by Gaia-like observatories or ground based experiments as 
the VLTI/GRAVITY instrument.

\section{Basics of astrometric microlensing}
During a microlensing event it is well known (see, e.g., Ref. \refcite{falco1992}) that multiple images form. In the case of a point-like object
lensing a background star, we indicate with $\mu_+$ and $\mu_-$ the magnifications (depending on the source-lens impact parameter) 
associated to the brighter and fainter images, respectively. 
Since the source moves in the lens plane with (transverse) velocity $v_{\perp}$, its projected coordinates (in units of the Einstein $R_E$ radius\footnote{The Einstein radius $R_E$ is given by
\begin{equation}
R_E \simeq D_L \theta_E
\end{equation}
where
 \begin{equation}
\theta_E=\left(\frac{4GM}{c^2}\frac{D_{S}-D_L}{D_L D_S}\right)^{\frac{1}{2}},
\end{equation}
and $M$ is the mass of the lens, $D_{S}$ and $D_L$ the distances from the observer to the source and lens, respectively.
} and with respect to a reference centered on the lens) are $\xi(t)=(t-t_0)/t_E$ and $\eta(t)=u_0$.

As a consequence, the centroid of the image pair (defined as the average position of the $+$ and $-$ images weighted 
with the associated magnification, Ref. \refcite{walker1995}) is 
\begin{equation}
\bar{u}\equiv\frac{\tilde{u}_+\mu_+ +\tilde{u}_-\mu_-}{\mu_+ +\mu_-}=\frac{u(u^2+3)}{u^2+2}.
\label{centroidpair}
\end{equation}
The astrophysical observable is the displacement vector of the combined image (the centroid) with respect to the source, i.e.
\begin{equation}
{\bf\Delta} \equiv{\bf \bar{u}}-{\bf u}=\frac{{\bf u}}{2+u^2}.
\label{shiftmodulus}
\end{equation}
with components (depending on time) given by\footnote{Note that all the previous angular distances are given in units of the Einstein angle $\theta_E$ which
sets the scale of the astrometric phenomenon.}
$\Delta_{\xi}=\xi(t)/(2+u^2)$ and $\Delta_{\eta}=u_0/(2+u^2)$.
Note that while the $\Delta_{\eta}$ component is symmetric with respect to $t_0$ and always positive,
the $\Delta_{\xi}$ component is an anti-symmetric function with minimum and maximum values occurring 
at $t_0 \pm t_E \sqrt{u_0^2+2}$, respectively\footnote{One can also verify that, in contrast to the magnification $\mu$ 
(which diverges for $u_0 \rightarrow 0$), the centroid shift assumes the 
maximum value equal to $\sqrt{2}/4$ for $u_0=\sqrt{2}$.
In particular, due to the anti-symmetry of the $\xi$ component, for $u_0<\sqrt{2}$ the shift goes 
through a minimum at $t=t_0$ and has two maxima at $t_0\pm t_E\sqrt{2-u_0^2}$. Conversely, 
for $u_0 \ge \sqrt{2}$, $\Delta$ assumes the single maximum 
value equal to $u_0/(u_0^2+2)$ at $t=t_0$.}.

\begin{figure}
\centering
{\includegraphics[width=6cm]{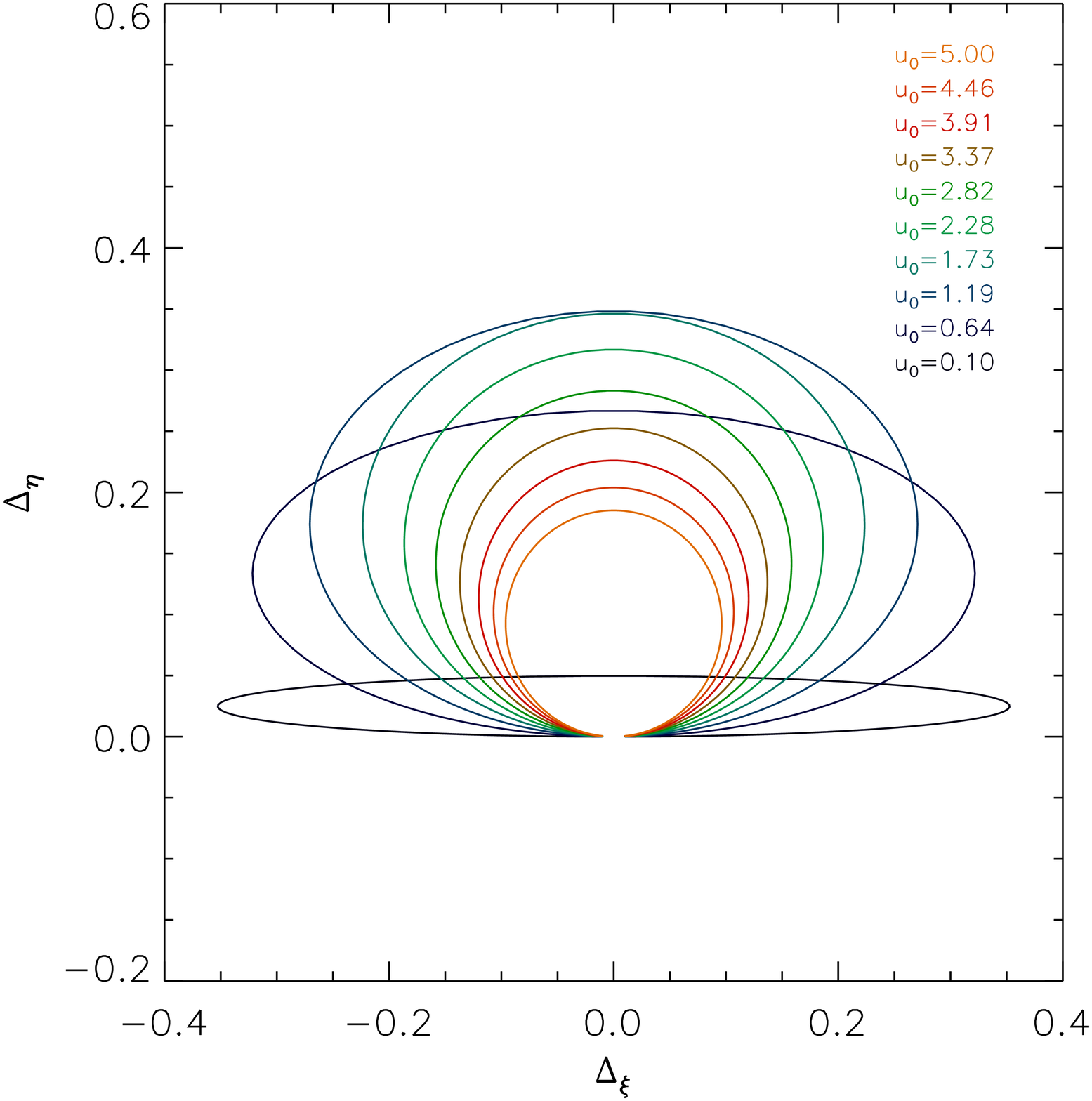}}
\caption{The centroid shift ellipse for different values of the impact parameter (see text for details).}
\label{fig5}
\end{figure}

As noted by Ref. \refcite{walker1995}, the centroid shift $\Delta$ traces (in the 
$\Delta_{\xi}, \Delta_{\eta}$ plane) an ellipse centered in the point 
$(0,b)$. This ellipse has a semi-major axis $a$ (along $\Delta_{\eta}$) and 
semi-minor axis $b$ (along $\Delta_{\xi}$) are
\begin{equation}
a=\frac{1}{2}\frac{1}{\sqrt{u_0^2+2}},~~~b=\frac{1}{2}\frac{u_0}{u_0^2+2}.
\label{axes}
\end{equation}
In Fig. \ref{fig5}, we give the centroid shift ellipses\footnote{Note also that from eq. (\ref{axes}) one easily finds that
\begin{equation}
u_0^2=2 (b/a)^2 \left[1-(b/a)^2\right]^{-1},
\label{u0fromaxes}
\end{equation}
which offers a way to measure the impact parameter directly.} for different $u_0$ values. Note that for $u_0 \rightarrow \infty$ the 
ellipse becomes a circle with radius $1/(2u_0)$ and degenerates into a straight line of length $1/\sqrt{2}$ when $u_0$ goes to zero. It is also obvious that
astrometric microlensing, being much more sensible to large impact parameters than the usual photometric microlensing, offers the possibility 
to predict close encounters\refcite{proximacentauri}. 


\subsection{Blending effect}
\label{blendingsinglelens}

\begin{figure}
\centering
{\includegraphics[width=6cm]{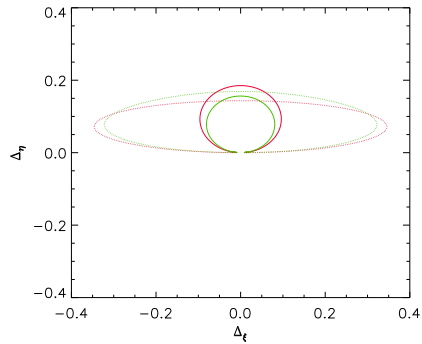}}
\caption{The centroid shift ellipses for a dark lens (red lines) and a luminous lens (green lines, for $f_L=0.2$) assuming different impact parameters are shown.
In particular, the solid lines are  obtained for $u_0=5$, while the dotted ones are for $u_0=0.3$.}
\label{blendedellipse}
\end{figure}

Generally speaking, in a microlensing event the intrinsic luminosity of the lens (or nearby stars) cannot be neglected. 
This effect, known as {\it blending}, represents  the fraction of light that does not get amplified 
but contributes to the photons collected during the observation. Blending influences in a trivial way also the astrometric measurements.
In fact, following Ref. \refcite{dominik2000}, we consider a luminous lens which is not resolved from the 
background source and define $f_l=L_L/L_S$ as the ratio between the lens and source luminosities. It is easy to see that the 
centroid position -- defined in eq. (\ref{centroidpair}) through a weighted average -- depends additionally on the new parameter $f_L$ as
\begin{equation}
\bar{u}=\frac{\tilde{u}_+\mu_+ +\tilde{u}_-\mu_-+\tilde{u}_L f_L}{\mu_+ +\mu_- + f_L},
\label{centroidpairblended}
\end{equation}
where $\tilde{u}_L$ is, in general, the position of the lens. Therefore, the centroid shift with respect to the source at rest is
\begin{equation}
\Delta_S\equiv\bar{u}-u=\frac{u-f_Lu^2\sqrt{4+u^2}}{2+u^2+f_Lu\sqrt{4+u^2}},
\label{shiftmodulusblended}
\end{equation}
since the used frame of reference is centered on the lens, i.e. $\tilde{u}_L=0$.
However, it is necessary to further subtract the proper motion of the apparent source object which corresponds to the superposition of the source and the luminous lens. In this case, the resulting (blended) centroid shift is
\begin{equation}
\Delta=\Delta_S+\frac{f_L}{1+f_L}u.
\label{shiftmodulusblended2}
\end{equation}
In analogy to the dark lens case, we define the blended shift components along the $\xi$ and $\eta$ axes as
\begin{equation}
\Delta_{\xi}=\Delta \cos{\alpha},~~~\Delta_{\eta}=\Delta \sin{\alpha},
\end{equation}
where
\begin{equation}
\alpha= \tan^{-1}\left(\frac{u_0 t_E}{\displaystyle{t-t_0}}\right).
\end{equation}
Note that for $f_L=0$ the result in eq. (\ref{shiftmodulusblended2}) 
reduces to that in eq. (\ref{centroidpair}). Furthermore, for $u\ll \sqrt{2}$ 
the blended centroid shift tends to zero linearly 
as 
\begin{equation}
\Delta\simeq \frac{u}{2}\left(\frac{1+3f_L}{1+f_L}\right),
\end{equation}
thus being enhanced by the factor $(1+3f_L)/(1+f_L)$. On the other hand, it goes as 
\begin{equation}
\Delta\simeq \frac{1}{(1+f_L)u}
\end{equation}
for $u\gg\sqrt{2}$, being reduced by a factor $1+f_L$ with respect to the dark lens case \cite{dominik2000} . 
In Fig. \ref{blendedellipse}, we plot the centroid shift ellipses for 
a dark lens (red lines) and a luminous 
lens (green lines, for $f_L=0.2$) assuming different impact parameter values.
In particular, solid lines are  obtained for $u_0=5$, while dotted 
lines are for $u_0=0.3$. It is then clear that in astrometric microlensing 
observations the luminosity of the lens cannot be in principle neglected
as (depending on $f_L$) it strongly affects the position of the image centroid.
In accordance to Ref. \refcite{lee2010}, when the lens becomes brighter, the trajectory becomes 
smaller and rounder and, for $u_0 \gg \sqrt{2}$, the blended centroid shift gets reduced 
with respect to the dark lens case. On the other hand, the blended shift $\Delta$ 
is sligthly larger than that 
in the dark lens case for $u_0 \ll \sqrt{2}$. As consequence of this fact, 
the ellipse like trajectory gets deformed so that the true value of $u_0$ can not 
be readily obtained.
%
\subsection{Finite-size source and single lens}
As observed by Ref. \refcite{walker1995}, the centroid shift trajectory 
for a finite-size source becomes more complicated with respect to the 
point-like star described above.
If one considers a source star with radius $\rho$, the centroid 
shift modulus can be evaluated
from eq. (\ref{centroidpair}) integrating both the numerator and 
denominator over the area subtended by the source and weighting with 
an appropriate function $S$, i.e.
\begin{equation}
\Delta=\frac{ \int_{{\rm A_{source}}}(\tilde{u}_+\mu_+ +\tilde{u}_-\mu_-) S dA}{ \int_{{\rm A_{source}}}(\mu_+ + \mu_-) S dA}-u.
\label{averagecentroid}
\end{equation}
Here, the factor $S$ accounts for the surface luminosity function of the star (see, e.g., Refs. \refcite{milne,afonso2000}, and \refcite{yoo2004}). 
Equivalently, it is straightforward to show that, in analogy to the point-like case described above, 
the centroid shift components along the $\xi$ and $\eta$ axes are given by
\begin{equation}
\Delta_{\xi}=\frac{ \int_{{\rm A_{source}}}(\tilde{u}_+\mu_+ +\tilde{u}_-\mu_-) \cos \alpha S dA}{ \int_{{\rm A_{source}}}(\mu_+ + \mu_-) S dA}-\frac{t-t_0}{t_E},
\label{averagecentroidx}
\end{equation}
and
\begin{equation}
\Delta_{\eta}=\frac{ \int_{{\rm A_{source}}}(\tilde{u}_+\mu_+ +\tilde{u}_-\mu_-) \sin \alpha S dA}{ \int_{{\rm A_{source}}}(\mu_+ + \mu_-) S dA}-u_0,
\label{averagecentroidy}
\end{equation}
respectively\footnote{Here, the angle $\alpha$ can be evaluated at any time $t$ (and 
for each source element with polar coordinates $(r,\theta)$ with respect to the source center, 
being $r$ and $\theta$ varying in the ranges $[0,\rho]$ and $[0,2\pi]$, respectively) 
\begin{equation}
\alpha= \displaystyle{\tan^{-1}\left(\frac{u_0+r\sin\theta}{\displaystyle{\frac{t-t_0}{t_E}}+r\cos \theta}\right)}.
\end{equation}
} . 
Then, the surface integrals in eqs. (\ref{averagecentroid})-(\ref{averagecentroidy}) can be solved numerically\footnote{Ref. \refcite{lee2010} solved the same problem by following the lens-centered coordinates approach
(see also Ref. \refcite{lee2009}) and, in accordance with our results, found that the point-like source approximation gives an 
over estimation of the astrometric signal whit respect to the finite source case. The same authors also applied the method to finite lenses, observing that the discontinuous astrometric trajectories obtained
by Ref. \refcite{takahashi} in the point-like approximation become continuous.}. We adopted the Vegas algorithm \cite{press} which is so powerful and robust to allow one to consider peculiar surface brightness profiles
(as in sources with stellar spots, see e.g. Refs. \refcite{giordano2015} and
\refcite{sajadian2015a}). Note that, for a uniformly bright circular source ($S=1$) of radius $\rho\ll u_0$, one recovers the approximated 
relation given by Ref. \refcite{walker1995}
\begin{equation}
\Delta\simeq \frac{u^3+3u}{u^2+2}\left[1+\frac{\rho^2(u^6+9u^4-6u^2-24)}{8u^2(u^2+2)(u^2+3)(u^2+4)}\right],
\label{approx}
\end{equation}
which holds only at the lowest order in $\rho/u$.
\begin{figure}
     \begin{center}
        \subfigure[]{%
            \includegraphics[width=0.4\textwidth]{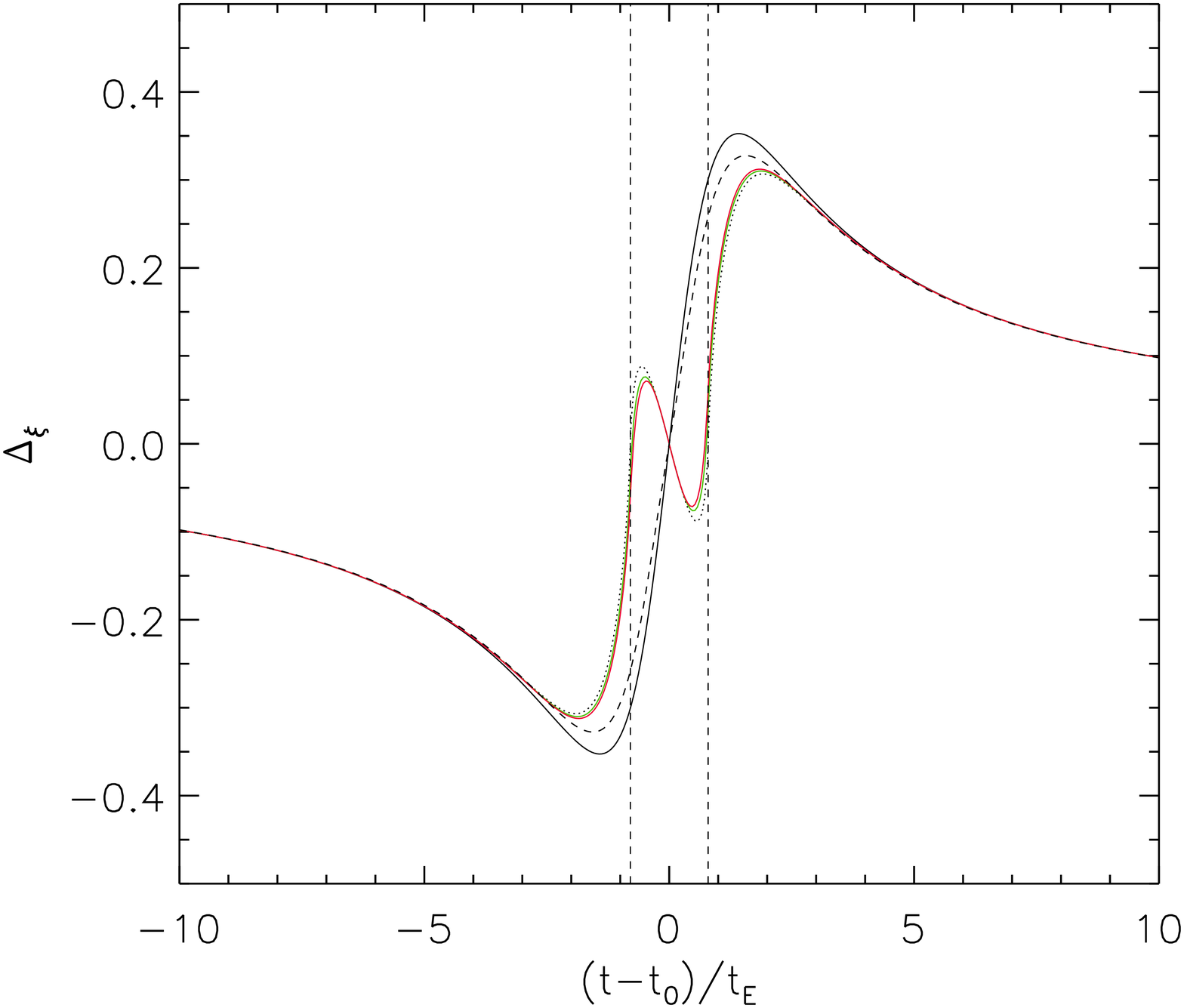}
                     }
        \subfigure[]{%
           \includegraphics[width=0.4\textwidth]{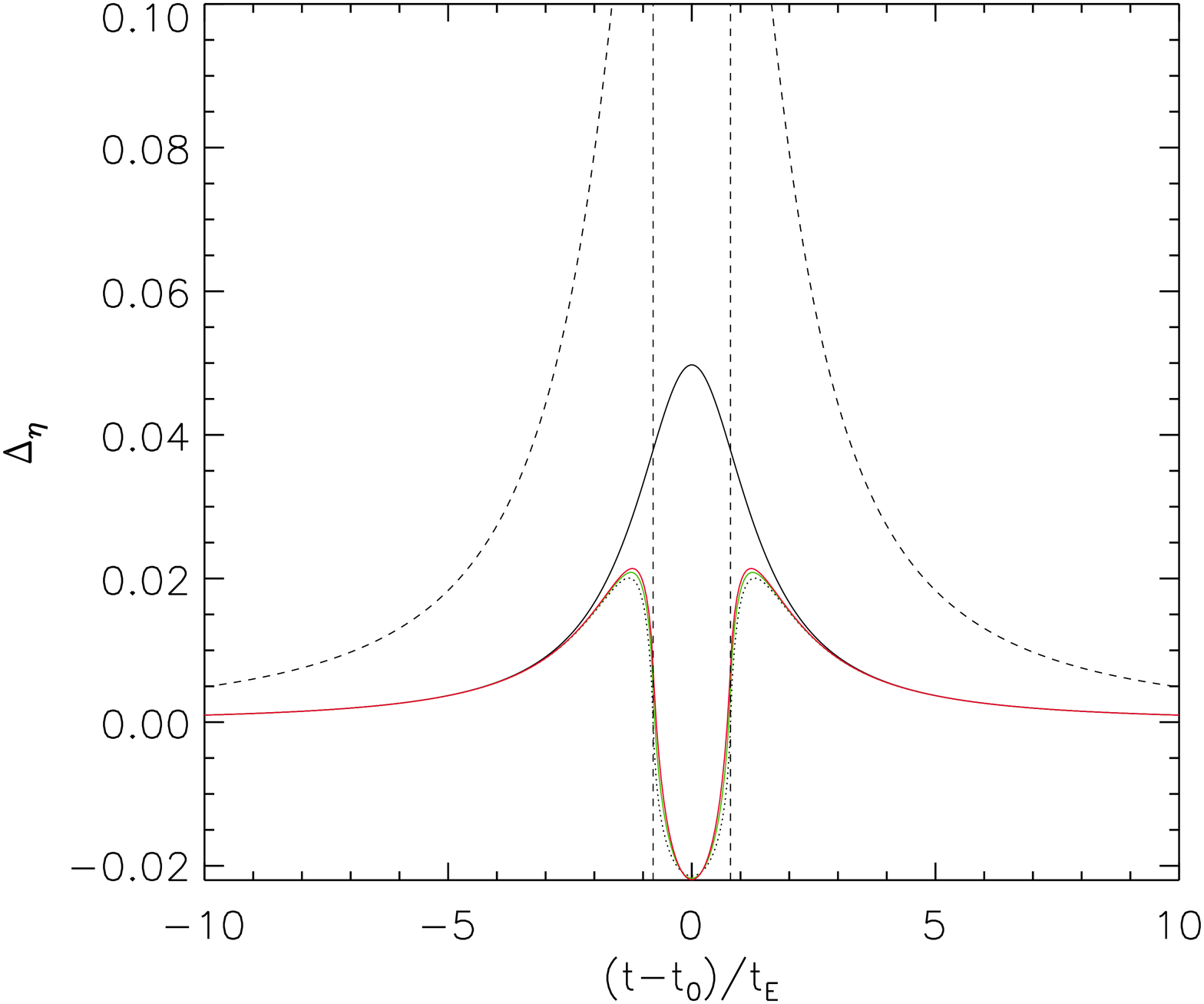}
                     }

\end{center}
 \caption{We show the $\xi$ (panel a) and the $\eta$ (panel b) components as a function 
of $(t-t_0)/t_E$ for point-like (black lines) and finite sources 
(red and green lines). See text for details)}
   \label{fig66162}
\end{figure}
In Figs. \ref{fig66162}, assuming $u_0=0.1$ and $\rho=0.8$, we adopted the strategy described 
above and give the $\xi$ (panel a) and $\eta$
(panel b) components of the centroid shift as a function of $(t-t_0)/t_E$ for a point-like source (black solid line),
a finite source uniformly bright (black dotted line) and for limb darkening profiles 
with $\Gamma=0.5$ (green solid line) and $\Gamma=0.8$ (red solid line).
For comparison, the black dashed curve has been obtained for $u_0=0.5$, $\rho=0.3$ and $\Gamma=0.5$.
The dashed vertical lines indicate the times at which $\rho=u$ (i.e. $(t-t_0)/t_E=\pm\sqrt{\rho^2-u_0^2}$). 
It goes without saying that, in the limit of small source-lens distance $u$ (and $u_0< \rho$), the
finite source effect drives the astrometric microlensing in a fashion completely different 
with respect to the point-like source case. In particular, a finite source gives 
rise to an astrometric microlensing shift smaller (within a few $t_E$) than that 
predicted for point-like stars and, more importantly, with a modified shape 
at least for small impact parameters. For example, in the case of the $\Delta_{\xi}$
component (panel a of Fig. \ref{fig66162}), there is only one maximum and one minimum at symmetric positions with respect to $t_0$ in the point-like case (black solid line), while additional
local extrema appear if one considers finite-size effects and small impact parameters (black dotted, red and green lines). 

Finally, as already observed in Ref. \refcite{lee2010}, when the impact parameter is smaller than the source radius,
the finite source effects introduce deformations and twists (the cloverleaf-like structures in Fig. \ref{fig63}) in the centroid shift trajectory in proximity of the closest approach. A similar behaviour is also found
in the case of the astrometry for binary lensing events when the binary components orbital motion is taken into account (see next Section). Here, black and green solid curves (obtained for $u_0=0.1$, $\rho=0.8$) represent the centroid shift ellipses for point-like source
and finite-size star ($\Gamma=0.5)$ in the $(\Delta_{\xi}, \Delta_{\eta})$ plane, respectively. The dashed trajectories (black line for point-like source, green one for an extend source with $\Gamma=0.5$) 
were obtained assuming $u_0=0.5$, and $\rho=0.3$. Note that with increasing impact parameter the finite-size effect introduces distortions at the distance of closest approach. 
\begin{figure}
\centering
{\includegraphics[width=8cm]{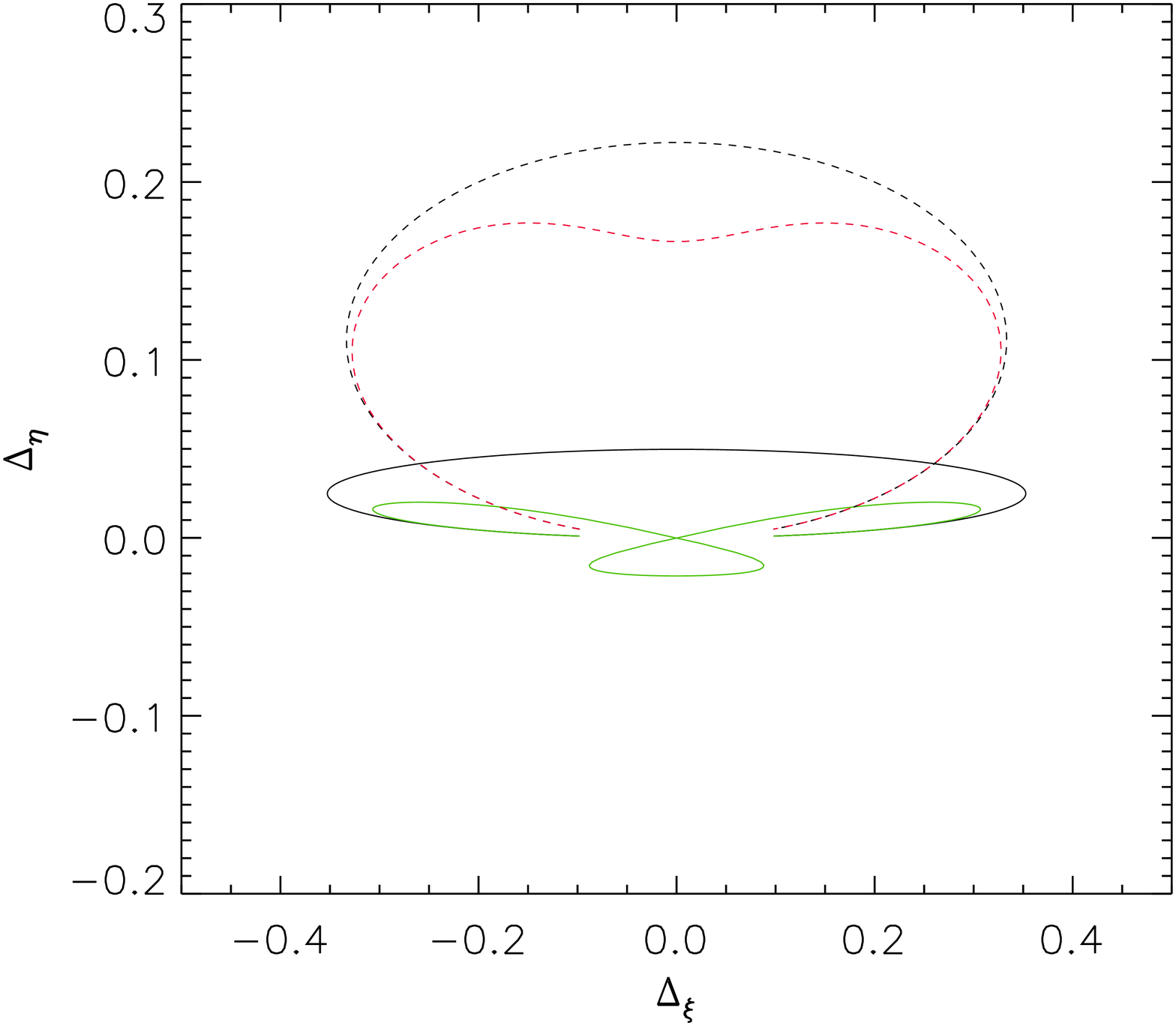}}
\caption{For an impact parameter smaller than the stellar radius, the centroid shift trajectory shows
the cloverleaf-like structures at the distance of the closest approach (see text for details).}
\label{fig63}
\end{figure}

The formalism introduced for the blending effect can be applied to account for luminous (single) lenses also in the case of finite-size sources. 
In fact, when considering a luminous lens, it is sufficient to replace the term $\mu_+ +\mu_-$ in the denominator 
of eq. (\ref{averagecentroid}) -- or equivalently in eqs. (\ref{averagecentroidx}) and (\ref{averagecentroidy}) -- with
$\mu_+ +\mu_- +f_L$. Finally, after performing the integrations and summing the term $f_L/(1+f_L)$ to account for the proper motion of the 
apparent source object, 
one is left with the centroid shift for a single luminous lens and a finite-source. Again, as expected, 
the blended centroid shift modulus changes 
(for a given source-lens distance) with respect to the case of the dark lens analogously to what seen for the blended lens case. 

\section{Binary lenses}
Approximately $50\%$ of all the stars are binary or multiple systems so that they offer an interesting channel for astrometric microlensing searches. 
The astrometric properties of gravitational microlensing caused by binary lenses were studied
by Refs. \refcite{griest1998} and \refcite{han1999}. Following Refs. \refcite{witt1990} and  \refcite{witt1995}, the lens equation of a binary
lens event in complex notation is given by
\begin{equation}
\zeta=z+\frac{m_1}{z_1+\bar{z}}+\frac{m_2}{z_2-\bar{z}},
\label{lensbinary1}
\end{equation}
where $m_1$ and $m_2$ are the masses of the two components (with $m_2<m_1$ so that $q=m_2/m_1<1$), $z_1$ and $z_2$ the positions
of the lenses (separated by $b$), and $\zeta=\xi+i\eta$ and $z=x+iy$ the positions of the source and images, respectively. Here, all lenghts are normalized to the Einstein ring size associated to the
total mass $M=m_1+m_2$ and the {\it bar} indicates the complex conjugate operation. Furthermore, we require that the lens components are located on the $\xi$ axis with the primary at $(-b/2,0)$ and the secondary at $(+b/2,0)$. 
The amplification $\mu_i$ of each image is obtained by the Jacobian ($J$) of the transformation in 
eq. (\ref{lensbinary1}) at the image position
with the image and source positions corresponding to infinite amplification (${\rm det}~J=0$) forming closed paths named critical (in the image plane) and caustic curves (in the lens plane), respectively.
The solution of the lens equation can be obtained either analytically (by solving an equivalent 5th order complex polinomial, 
see Ref. \refcite{witt1990,sgnumerical}) or numerically with a inverse
ray-trace algorithm as described in Ref. \refcite{falco1992} in
which one simply shoots many photons from the observer back to the lens plane and, through the lens equation,
recording those falling onto the source. Finally, the image amplification\footnote{Depending on the binary lens parameters and on the source path (a straight line forming an angle $\theta$ with respect to the binary lens axis), the procedure
results in a variety of microlensing light curves dramatically different from a typical Paczinsky light curves. For a complete description of the possible caustic
shapes and classification of the binary lens microlensing light curve, we address the reader to Refs. \refcite{Pejcha2009} and \refcite{liebig}.
} can be thereby obtained as it is proportional to
the number of photons collected at a given point. Here, we adopt a hybrid method consisting in using the robust inverse ray-trace
method close to the caustics and solving the 5th order complex polynomial far from them (see e.g. Refs. \refcite{nucita2014,giordano2015}). Ref. \refcite{han1999} found that the position of the source star centroid is the average of the positions of the individual images weighted by each amplification $\mu_i$, i.e.
\begin{equation}
(\xi_c,\eta_c)=\left(\sum_i \mu_i x_i/\mu, \sum_i \mu_i y_i/\mu\right),
\label{binaryastrometry}
\end{equation}
where $\mu$ is the total amplification, i.e. $\mu=\sum_i\mu_i$ and $i$ runs over the image number. Finally, the centroid shift with respect 
to the position of the unlensed star has components
\begin{equation}
(\Delta_{\xi},\Delta_{\eta})=(\xi_c-\xi, \eta_c-\eta),
\label{binaryastrometryeffective}
\end{equation}
which, of course, depend on the time $t$ since the source is moving in the lens plane. By investigating this issue,
Ref. \refcite{han1999} found dramatic changes of the astrometric shift trajectories from the ellipse path typical of a single lens event. In fact,
the shift trajectories associated to binary lenses are characterized by distortions, twists and big jumps depending on the lens parameters (separation and mass ratio) and source path.
In particular, it was shown that distortions and twistings appear for non caustic crossing events, with small deformations of the shift ellipse for $b\ll u_0$ and loops at
the closest approach for  $b\sim u_0$. Conversely, when the source path crosses over the caustics (and this has a much larger probability to occur for large values of $b$, and $q$ and finite-size sources) 
the shift trajectories manifest with big jumps. 
Furthermore, it has been also shown that the degeneracy that affects the photometric microlensing
can be solved when the astrometry is taken into account as the shift trajectories strongly depend on the microlensing parameters.
However, as it will be clear in the following, this conclusion must be relaxed and taken with caution when additional effects are considered.
Note that the approach described here is general as one can treat a binary lens system with any value of mass ratio and separation. Furthermore, 
these results are in agreement with that derived by \cite{sajadian2014} (who describes the astrometric microlensing with rotating stellar-mass black holes)
in the small $q$ limit and large separation $b$. The benefits of the above described method is that it can account for any binary system regardless the values of the interesting parameters and without any approximation

Recent theoretical analyses (see e.g. Refs. \refcite{dominik1997,penny2011a,penny2011b,nucita2014,sajadian2014,giordano2015})
and microlensing observations (see e.g. Refs. \refcite{park2015,Skowron2015,udalski2015}) showed that, in some cases, the orbital
motion of the binary lens system cannot be neglected. This motion can be accounted for by solving the associated Kepler problem in a closed way. Alternatively,
with a simpler and approximated approach, it is possible to parametrize the motion of the projected lenses. In this case, it is only necessary to know the rates of change in time
of the projected binary lens separation ($db/dt$) and of its orientation ($d\alpha/dt$) with respect to a fixed axis. However, here we do not consider the first quantity which can
be easily ignored as it represents a second order effect for binary lenses with large inclination angles.
By recognizing that $d\alpha/dt=-d\theta/dt$, this results in a source path which is curved in the lens plane. 
In order to get easily any difference, we further required that the source paths coincide at the distance of closest approach.
\begin{figure*}
\centering
{\includegraphics[width=10cm]{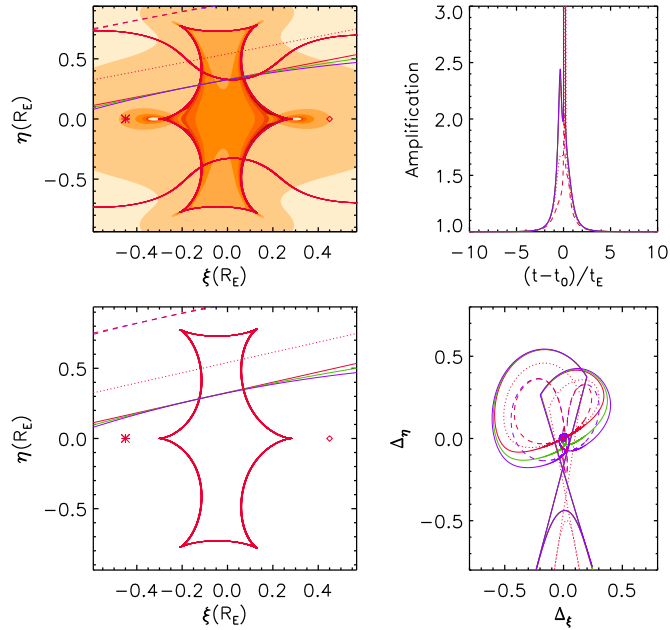}}
\caption{Upper-left panel: the amplification map with superimposed the critic and  caustic curves and source paths. Bottom-left panel: a zoom around the caustic curves. Upper-right panel: the event light curve. Bottom-right panel:
the ellipse-like trajectories of the centroid shift. The lens is a binary system with parameters $b=q=0.9$. See text for the meaning of the different line styles and colors.}
\label{fig9}
\end{figure*}
As an example, in Fig. \ref{fig9} we give (from the upper-left panel and in counterclock-wise order) the amplification map with superimposed the critic, caustic curves and source path,
a zoom around the caustic curves, the ellipse-like trajectories of the centroid shift, and the resulting amplification light curve.
In the panels, the primary star and its companion are indicated by an asterisk  and a diamond, respectively. The shift trajectories for the particular binary
lens ($b=q=0.9$) are obtained for several values of the impact parameter. In particular, we fixed $u_0=0.3$ (solid lines),
$u_0=0.5$ (dotted lines), and  $u_0=0.9$ (dashed lines). Here, red color is used for the static binary lens while green and blue ones are used for rotating binary with orbital
periods of $100t_E$ and $50t_E$, respectively. In the case of a rotating binary lens, we found that the shift trajectories differ (around the distance of minimum approach) from those expected in the static 
case and, depending on the impact parameter $u_0$, multiple twists appear after the microlensing event had (photometrically) finished. 

The finitess of the source star can be accounted for by integrating the numerators and denominators on the right-hand side of eq. (\ref{binaryastrometry}) over the projected surface of the star and considering the appropiate 
limb-darkening profile in analogy to the single lens case. 
Furthermore, also in the case of a binary lens, the intrinsic luminosity of the primary and/or secondary components can strongly affect the astrometric signal. After defining 
the intrinsic luminosity ratios of the primary and secondary lens as $f_1=L_1/L_S$, and $f_2=L_2/L_S$, respectively, we followed the same procedure
described in Section \ref{blendingsinglelens}. Taking into account the luminosity and position of the two lenses\footnote{In the adopted frame of reference, the primary star 
is located on the $\xi$ axis at coordinates $(-d, 0)$ while its companion is in $(d,0)$, where $d=b/2$.}, eq. (\ref{binaryastrometry}) becomes
\begin{equation}
(\xi_c,\eta_c)=\left[\frac{\left(\sum_i \mu_i x_i-f_1d+f_2d\right)}{\left(\mu+f_1+f_2\right)}, \frac{\left(\sum_i \mu_i y_i\right)}{\left(\mu +f_1+f_2\right)}\right].
\label{binaryastrometryblended}
\end{equation}
Finally, when accounting for the position of the apparent source object, we obtain that the centroid shift components are
\begin{equation}
(\Delta_{\xi},\Delta_{\eta})=\left[\xi_c-\xi+\frac{f_1(\xi+d)+f_2(\xi-d)}{1+f_1+f_2}, \eta_c-\eta+\frac{(f_1+f_2)\eta}{1+f_1+f_2}\right],
\label{binaryastrometryeffectiveblended}
\end{equation}
which can be further simplified for a planetary system by assuming $f_2=0$. As expected, the astrometric signal is quite different from that derived in absence of the blending effect. 
Obviously, these simulations clearly show that, depending on the event parameters (binary separation and mass ratio) and source/lens characteristics (finite size, intrinsic lens luminosity and binary orbital motion), 
all the effects are of the same order. In real data analysis, only a very accurate modelling that accounts for these effects would allow one to discriminate among different
scenario and get information about the physical properties involved.


\section{Binary sources}
Identifying if a lens is constituted by a binary object is, generally, quiet easy especially in microlensing events  characterized  by  caustic  crossings as the  resulting  light  curve
shows strong deviations with respect to a classical Paczynski profile. 
On the contrary, binary sources lensed by a point like objects exhibits amplification curves with only minor anomalies thus making hard, if not impossible, 
to identify the source binarity.  In this respect, as recently showed by Ref. \refcite{nucita2016}, astrometric microlensing offers a new way to discover these kind of signatures.
The total centroid shift at time $t$ can be obtained via a weighted 
average on the individual source component amplifications and using as reference position the centre of light between the unlensed source components, i.e.\cite{han2001astrometry}  
\begin{equation}
{\bf \Delta _{bs}}=\frac{\mu_1F_1({\bf u_1}+{\bf \Delta_1})+\mu_2F_2({\bf u_2}+{\bf \Delta_2})}{\mu_1F_1+\mu_2F_2}-\frac{F_1{\bf u_1}+F_2{\bf u_2}}{F_1+F_2},
\label{deltatotalbs}
\end{equation}
where ${\bf u_i}$ are the distances between the lens and the individual binary source components, $\mu_i$ and ${\bf \Delta_i}$ the magnification factors and centroid shifts of the 
two single sources (see Section 1) having luminosity $F_i$ with subscripts $i=1$ and $i=2$ for the primary object and its companion, respectively. 
As already discussed in the previous Section the orbital motion may play a crucial role. Hence, as in Ref. \refcite{nucita2016} one can determine the astrometric shift components by requiring that the two sources move around 
the common center of mass.

\begin{figure*}
     \begin{center}
        \subfigure[]{%
           \includegraphics[width=0.4\textwidth]{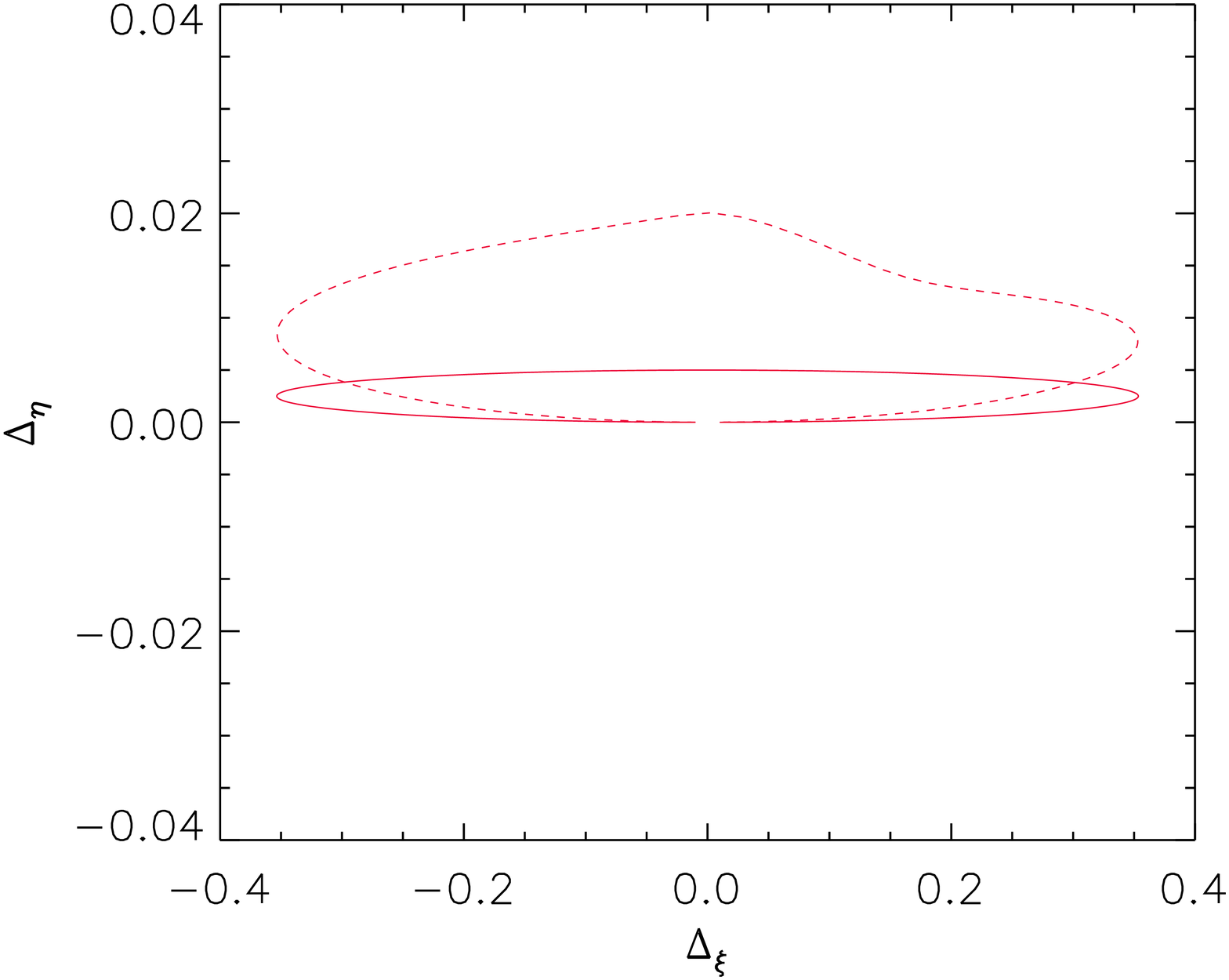}
        }
\subfigure[]{%
            \includegraphics[width=0.4\textwidth]{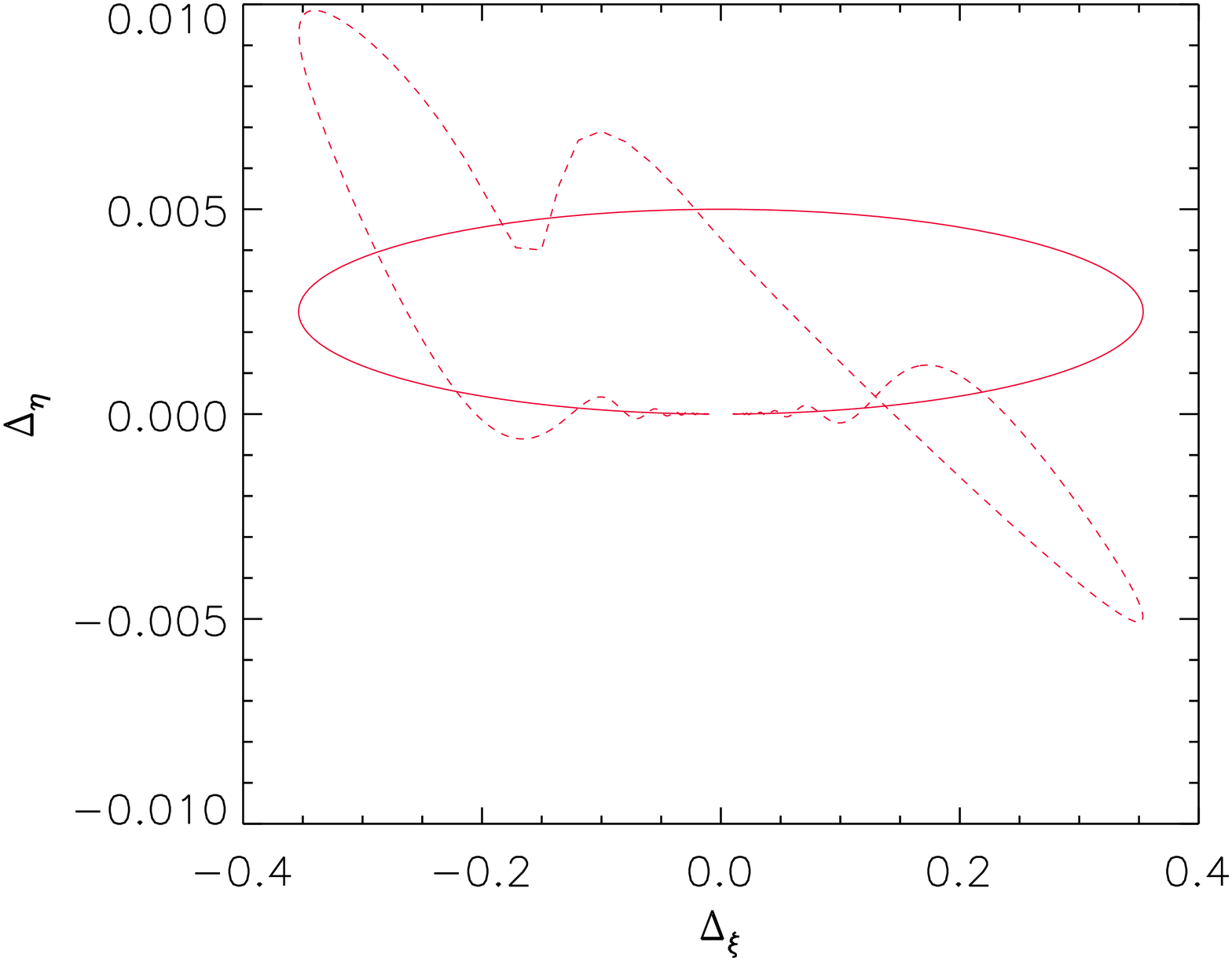}
        }
    \end{center}
 \caption{The strometric shift  (dashed curve) for a binary source static (left panel) and rotating (right panel). The lens is a
$1$ M$_{\odot}$ object located at $D_L=1$ kpc and moving with transverse velocity $100$ km s$^{-1}$ so that the Einstein ring is 2.7 mas.}
   \label{fig2paper2}
\end{figure*}

As an example, in Figure \ref{fig2paper2}, we consider the expected astrometric microlensing signal for a static  (panel a) and rotating (panels c) binary source, respectively. Here, we assumed two objects 
with masses $m_1=1$ M$_{\odot}$, and $m_2=0.1$ M$_{\odot}$, separated by $1$ AU and fixed the intrinsic luminosities to $F_1=1$ L$_{\odot}$, and $F_2=0.01$ L$_{\odot}$. We furthermore set $u_0=0.01$. 
For such case, the binary source orbital period turns out to be $P\simeq 370$ days.
Here, the solid curve represents the centroid shift ellipse\footnote{Note that, for the simulated cases, being $\theta_E\simeq 2.7$ mas, the astrometric signal results well within the astrometric 
precision of the Gaia satellite in five years of integration. This opens the possibility to detect binary systems as sources 
of astrometric microlensing events and characterize their physical parameters (mass ratio, projected separation and orbital period).} expected for a single source located in the center 
of mass of the binary source system. As it is clear, the presence of a binary source system\footnote{  
Gaia-like observatories might also detect astrometric microlensing 
events involving both binary sources and binary lenses. For such cases, eq. (\ref{deltatotalbs}) continues to remain valid provided that the centroid shifts ${\rm \Delta_i}$ of 
each components of the binary source system are obtained solving numerically the two body lens equation. Hence, eq. (\ref{binaryastrometryeffective}) can be applied for each of 
the sources obtaining (see also Ref. \refcite{han2001astrometry}) $(\Delta_{\xi,i}, \Delta_{\eta,i})=(\xi_{c,i}-\xi_i, \eta_{c,i}-\eta_i)$, 
where the positions of the source star centroid are simply the average of the locations of the individual images wighted by each amplification $\mu_{j,i}$, i.e.
\begin{equation}
(\xi_{c,i}, \eta_{c,i})=\left(\sum_j\mu_{j,i}x_i/\mu_i, \sum_j\mu_{j,i}y_i/\mu_i\right).
\end{equation}
Here, $\mu_i$ is the total amplification (i.e. $\mu_i=\sum_j \mu_{j,i}$, with $j$ running over the image number) and, as above, $i=1,2$ indicates the primary and secondary component of the binary source system.}
introduces deformations of the astrometric signal with respect to the pure ellipse with the orbital motion introducing characteristic periodic features
\footnote{Orbital features are also expected when the Earth parallax is taken into account. In this case, one could 
use the formalism provided by Ref. \refcite{dominik1997} in the approximation for small orbital eccentricity and easily gets
the expected astrometric curves for long duration events (see, e.g., Ref. \refcite{nucita2016}).} .

\section{Astrometric microlensing by black holes}
Recently, the challenging possibility to detect intermediate 
mass black holes\footnote{Based on the extapolation of the black hole-to-bulge mass relation \cite{magorrian}, one expects to find IMBHs with mass $\simeq 10^4-10^5$ M$_{\odot}$ in globular clusters and 
close dwarf galaxies \cite{maccarone2005}. For recent studies on IMBHs expected to be host in such stellar systems we address the reader to Refs.
\refcite{reines2011,nucitaimbh2013a,nucitaimbh2013b,manni2015,reines2015,nucitaimbh2016}.} (IMBHs) in globular clusters via astrometric microlensing has been investigated\cite{sahu2016}. 
By selecting clusters close to the line of sight to the Galactic Bulge and the Small Magellanic Cloud, Ref. \refcite{sahu2016} performed accurate simulations in order to estimate the probabilities of detecting 
the astrometric signatures caused by black hole lensing and found that, already with archival Hubble Space Telescope data, the chance to get such an event is not negligible. 
As an example, for the M22 globular cluster it is found that a central IMBH with mass $\simeq 10^5$ M$_{\odot}$ would induce an astrometry signal detectable over a background star with a probability 
of $\simeq 86\%$. As a matter of fact, a bulge star lensed by such IMBH (at the distance of $\simeq 4$ kpc) suffers of an astrometric signal of a few mill-arcseconds when the impact parameter (in units of the Einstein radius)  is as large as 100, i.e. well within the detection capabilities of present instrumentation (see, e.g., Ref. \refcite{eyer}).

\section*{Acknowledgments}
{We acknowledge the support by the INFN projects TAsP and Euclid.}

 \end{document}